\documentclass[%
twocolumn,
groupedaddress,
 amsmath,amssymb,
 aps,
pra,
showpacs,
floatfix,
longbibliography,
]{revtex4-1}

\usepackage{graphicx}
\usepackage{dcolumn}
\usepackage{bm}

\AtBeginDocument{%
	\newwrite\bibnotes
	\def\bibnotesext{Notes.bib}
	\immediate\openout\bibnotes=\jobname\bibnotesext
	\immediate\write\bibnotes{@CONTROL{REVTEX41Control}}
	\immediate\write\bibnotes{@CONTROL{%
			apsrev41Control,author="08",editor="1",pages="1",title="0",year="1"}}
	\if@filesw
	\immediate\write\@auxout{\string\citation{apsrev41Control}}%
	\fi
}%

\begin{document}

\preprint{APS/123-QED}

\title{Quench spot detection for superconducting accelerator cavities via flow visualization in superfluid helium-4}

\author{Shiran Bao${^{1,2}}$}
\author{Wei Guo${^{1,2,}}$}%
\email{Corresponding: wguo@magnet.fsu.edu}
\affiliation{${^{1}}$ National High Magnetic Field Laboratory,  1800 East Paul Dirac Drive, Tallahassee, Florida 32310, USA,\\
	${^{2}}$ Mechanical Engineering Department, Florida State University, Tallahassee, Florida 32310, USA}

\date{\today}

\begin{abstract}
Superconducting ratio-frequency (SRF) cavities, cooled by superfluid helium-4 (He II), are key components in modern particle accelerators. Quenches in SRF cavities caused by Joule heating from local surface defects can severely limit the maximum achievable accelerating field. Existing methods for quench spot detection include temperature mapping and second-sound triangulation. These methods are useful but all have known limitations. Here we describe a new method for surface quench spot detection by visualizing the heat transfer in He II via tracking He$_2^*$ molecular tracer lines. A proof-of-concept experiment has been conducted, in which a miniature heater mounted on a plate was pulsed on to simulate a surface quench spot. A He$_2^*$ tracer line created nearby the heater deforms due to the counterflow heat transfer in He II. By analyzing the tracer-line deformation, we can well reproduce the heater location within a few hundred microns, which clearly demonstrates the feasibility of this new technology. Our analysis also reveals that the heat content transported in He II is only a small fraction of the total input heat energy. We show that the remaining energy is essentially consumed in the formation of a cavitation zone near the heater. By estimating the size of this cavitation zone, we discuss how the existence of the cavitation zone may explain a decades-long puzzle observed in many second-sound triangulation experiments.

\end{abstract}

\maketitle


\section{\label{sec:introduction}Introduction}
Superconducting ratio-frequency (SRF) cavities are key components in many modern particle accelerators due to their high Q factors \cite{Nassiri2016History}. When these cavities are cooled by superfluid helium-4 (He II) to around 2 K, electric power injected in the cavities can generate extremely high electric field that allows charged particles to be accelerated to high energies over short distances. The maximum accelerating gradient of typical SRF cavities is in the range of 25-30 MW/m with a record value of 45 MW/m \cite{Padamsee201750}. This maximum gradient is limited by the breakdown of the superconductivity of the cavities, a phenomenon known as ``quench''. Quenches can be caused by Joule heating from tiny (i.e., about 1-10$^2$ $\mu$m in radius) resistive surface defects on the cavity inner walls (such as impurities, pits, cracks, scratches) or local phase transition caused by trapped magnetic fluxes \cite{padamsee_rf_2008}. When the temperature at the edge of the resistive region exceeds the superconducting phase transition temperature, surrounding region also becomes normal conducting. This process then spreads out rapidly over the entire cavity, causing the stored energy to convert to heat around the defect area within a few milliseconds \cite{padamsee_rf_2008}.

The maximum accelerating gradient of SRF cavities can be improved by removing the surface defects via mechanical grinding, tumbling the cavity, and electron or laser re-melting \cite{ge_repair_2011, watanabe_cavity_2011, Conway2017Instrumentation}. In order to locate the surface defects, a multi-channel temperature mapping (T-mapping) method was first developed \cite{Knobloch1994Design,canabal_full_2008}. This method requires a large number of temperature sensor (i.e., over 1000) in good thermal contact with the outer surface of the cavity and is often applied at a cavity accelerating gradient just below the quench threshold. The continuous Joule heating from a surface defect raises the local temperature, which manifests the defect location in the temperature map. Despite the usefulness of T-mapping, the spatial resolution is limited by the spacing between sensors (i.e., of order 1 cm). Furthermore, the installation of the large amount of sensors makes the application of this method an extremely laborious task \cite{Conway2017Instrumentation}. An alternative way to apply T-mapping is to scan the cavity surface using a rotating arm with just a few sensors arranged in a stripe. Nevertheless, to allow smooth rotation, a gap between the sensors and the cavity surface is required, which limits the detection sensitivity \cite{shu_novel_1996, sawamura_cavity_2008}.

A more convenient non-contacting quench spot detection method based on second-sound triangulation was later introduced by a team at Cornell University \cite{conway_defect_2010}. This method makes use of the unique properties of He II. In the superfluid phase below $T_\lambda$$\simeq$2.17 K, He II can be regarded as a mixture of two interpenetrating fluids: a viscous normal fluid that carries all the entropy and an inviscid superfluid that possesses zero entropy \cite{tilley_book}. This two-fluid system can support two distinct sound modes: the first sound, i.e., an ordinary pressure-density wave in which both fluids move in phase, and the second sound, i.e., a temperature-entropy wave in which the two fluids move out of phase \cite{landau_fluid_1987}. When the cavity quenches, the heat generated in the defect region is conducted promptly to the cavity outer surface. This heat is then released into He II, causing the generation and propagation of second-sound waves, in which a counterflow of the two fluids can establish \cite{VanSciver2012Helium}. These second-sound waves can be detected using sensors such as oscillating superleak transducers (OST) \cite{conway_defect_2010,Sherlock1970Oscillating}, resistive temperature detectors (RTD) \cite{Shepard1979Development}, and transition edge sensors (TES) \cite{Lunt2017Towards}. By measuring the time-of-arrival of the second-sound waves at three or more such sensors and implementing triangulation, the surface defect can in principle be located. However, a mystery was observed in many second-sound triangulation experiments. In order for the triangulation to converge to the cavity surface, a second-sound speed faster than tabulated values must be assumed \cite{conway_defect_2010, markham_quench_2015, eichhorn_mystery_2014, peters_advanced_2014, junginger_high_2015, plouin_experimental_2013}. This converged location can have an uncertainty of 5-10 mm from the actual defect location \cite{bertucci_quench_2013}, making it difficult for subsequent optical inspection of the sub-millimeter defect \cite{Wenskat2017Automated,Iwashita2008Development}. Various models have been proposed to explain the puzzling fast second sound, such as spreading of the heat in the cavity walls \cite{peters_advanced_2014, markham_quench_2015}, possible delay in detecting the start of the quench \cite{markham_quench_2015, eichhorn_mystery_2014}, and nonlinear effect that affects the second-sound shock speed at high heat fluxes \cite{Torczynski1984On, junginger_high_2015}. However, none of these models can offer a convincing explanation that systematically accounts for various observations \cite{Conway2017Instrumentation, junginger_high_2015, plouin_experimental_2013}.

In this paper, we discuss a new non-contacting method for quench spot detection by visualizing quench-induced thermal counterflow in He II using a He$_2^*$ molecular tracer-line tagging technique developed in our laboratory \cite{Guo2014Visualization, Gao2015Producing, Marakov2015Visualization}. A proof-of-concept experiment has been conducted, in which a miniature heater mounted on a plate in He II was utilized to simulate a surface quench spot. A He$_2^*$ tracer line created nearby the heater deforms as a result of the transient counterflow. The experimental techniques are discussed in detail in Sec.~\ref{sec:exp}. In Sec.~\ref{sec:result}, we present the analysis results. We show that by analyzing the tracer-line deformation, the heater location can be determined with an uncertainty of only a few hundred microns. Our analysis also reveals that the heat content transported in He II is just a small fraction of the total input heat energy. We show, in Sec.~\ref{sec:discussion}, that the remaining energy is largely consumed in the formation of a cavitation zone near the heater. The creation and collapsing of vapor bubbles inside the cavitation zone can convert the heat energy to acoustic energy. By estimating the size of the cavitation zone, we propose a model that explains the puzzling faster second sound and gives estimated excess second-sound velocity in quantitative agreement with the observations from previous triangulation experiments. We also discuss how our visualization technique may be advanced for practical SRF cavity quench spot detection. A brief summary is given in Section~\ref{sec:summary}.

\section{\label{sec:exp} Experimental techniques}
\begin{figure}
	\includegraphics[scale=0.44]{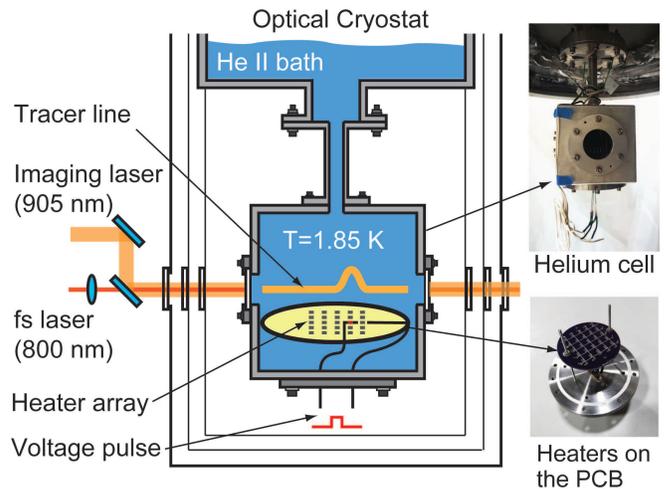}
	\caption{\label{fig:fig1} A schematic of the experimental setup (not to scale).}
\end{figure}

The optical cryostat used in our proof-of-concept experiment is shown schematically in Fig.~\ref{fig:fig1}. An aluminum cubic helium cell with an inner side length of 3 inches is connected to a pumped helium bath whose temperature was maintained at 1.85 K, typical of the operation temperature of SRF cavities, by regulating the vapor pressure in the bath. To examine the relevant heat transfer processes in a real quench event, we note that the energy stored in a SRF cavity (i.e., of order 1-10 J) is converted to heat in a few milliseconds \cite{padamsee_rf_2008}. As the heat is conducted to the cavity outer surface, the heated region can expand to an area of order 1-10 cm$^2$ with the hottest spot at the area center. Therefore, the instantaneous heat flux into He II is of order 10$^2$-10$^3$ W/cm$^2$ \cite{junginger_high_2015}. In order to simulate this heat flux in our experiment, we utilize an array of 5$\times$5 miniature thick-film resistors (surface area $A_h$=0.8$\times$0.8 mm$^2$) installed on an insulated Printed Circuit Board (PCB). The resistance of these resistor heaters was measured to be $49.7\pm0.3$ $\Omega$ at 1.85 K. A rectangular voltage pulse with a duration of 1-4 ms and an adjustable amplitude up to 10 V can be applied to a selected heater, giving rise to a heat flux into He II up to 315 W/cm$^2$ at the heater surface. The size of our heaters is relatively small compared to the size of the heated area on cavity outer surface. Nevertheless, this heater size is common among quench spot testing experiments and simulations \cite{Quadt2012Response,fouaidy_calibration_2013,fouaidy_detection_2017,junginger_high_2015} and is also desired for testing the resolution of our flow-visualization based detection method.

The heat ejected into He II leads to the generation of a second-sound shock wave followed by a thermal counterflow of the two fluids (see detailed discussions in Sec. \ref{sec:result}). In order to visualize the flow of the normal fluid that carries the heat content in He II, we adopted our He$_2^*$ molecular tracer-line tagging technique \cite{Gao2015Producing}. A 35-fs pulsed laser beam at 800 nm with a repetition rate of 5 kHz and a pulse energy of about 60 $\mu$J was focused to pass through the helium cell. Due to the strong instantaneous laser field, some helium atoms are ionized along the fs-laser beam path in the focal region. The recombination of the electrons and helium ions then leads to the formation of metastable He$_2^*$ triplet molecules \cite{Benderskii1999Photodynamics}. These molecules form tiny bubbles in He II (i.e., 6 {\AA} in radius \cite{Benderskii-JCP_2002}) and have a lifetime of about 13 s \cite{McKinsey-PRA_1999}. Above 1 K, they are completely entrained by the viscous normal fluid since Stokes drag easily dominates other forces for small molecules. This line of He$_2^*$ molecular tracers can then be driven to produce 640 nm fluorescent light by a 5-ns pulsed imaging laser at 905 nm \cite{Rellergert2008Detection}. The imaging laser in our experiment has a repetition rate of 500 Hz and is shaped into a laser sheet (thickness: 1 mm, height: 5 mm) that covers the entire region traversed by the tracer lines. The fluorescence is captured by an intensified CCD (ICCD) camera mounted perpendicular to the tracer-line plane. Typically, 5-6 imaging pulses are used to produce good quality images. This flow visualization technique has been successfully utilized in our quantitative studies of quantum turbulence in He II \cite{Marakov2015Visualization, Gao-PRB_2016, Gao-JETP_2016, Gao-PRB_2017, Gao2018Dissipation, Varga-PRF_2018}.

\begin{figure}
	\includegraphics[scale=0.44]{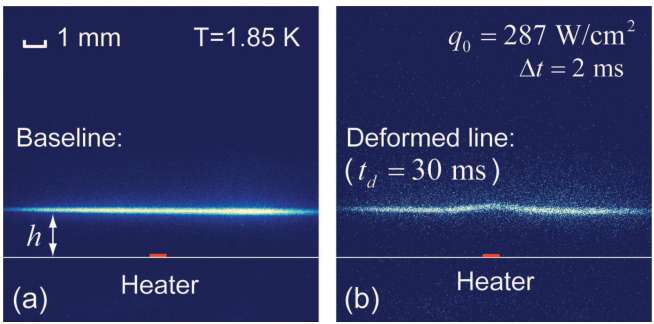}
	\caption{\label{fig:fig2} Typical images showing (a) a baseline created at $h$=2.13 mm above the heater, and (b) a deformed tracer line following a heat pulse of $q_0$=287 W/cm$^2$ and $\Delta{t}=2$ ms.}
\end{figure}

By adjusting the position of the fs-laser beam, we can create a He$_2^*$ tracer line nearly in parallel to the PCB at a height $h$ right above a chosen heater. Without turning on the heater, an image of the tracer line can be taken as a reference (i.e., the baseline). A typical baseline image is shown in Fig. \ref{fig:fig2} (a). The thickness of the baseline is about 100 $\mu$m, which matches the thickness of the fs-laser beam in the focal region. The time delay between the creation and imaging of the baseline (i.e., the drift time $t_d$) is set to zero. Indeed, when there is no flow in He II, the baseline remains straight at its original position regardless of the duration of the drift time. In the tests for locating the heater, we first create a tracer line and then turn on the heater by applying a voltage pulse of duration $\Delta{t}$. The instantaneous heating power $\dot{Q}_0$ (and hence the heat flux $q_0$=$\dot{Q}_0/A_h$) can be controlled by varying the voltage on the heater. An initially straight tracer line deforms due to the normal-fluid flow accompanying the heat transfer. After a typical drift time $t_d$ of 20-30 ms, we send in the imaging pulses to visualize the deformed line. Fig. \ref{fig:fig2} (b) shows an example image of the deformed tracer line following a heat pulse of $q_0$=287 W/cm$^2$ and $\Delta{t}$=2 ms. Note that the heat content is carried by second-sound waves in He II, which propagates at the second-sound speed, i.e., $c_2=19.5$ m/s at 1.85 K \cite{Donnelly-JPCRD_1998}. It takes less than 1 ms for the second-sound waves to pass across the entire tracer line. Indeed, when $t_d$ is greater than 2-3 milliseconds, we always observe nearly the same deformation of the tracer line.

\section{\label{sec:result} Analysis and results}
Apparently, the deformation of the tracer lines contains important information about the heater location and the heat content transported through He II. In order to extract this information, a detailed understanding of the relevant heat transfer processes in He II and the expected motion of the tracer lines is needed.

\begin{figure}[tbp]
	\includegraphics[scale=0.4]{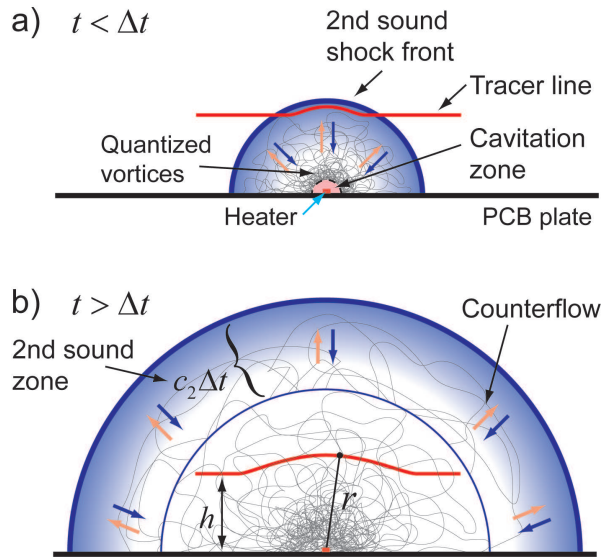}
	\caption{\label{fig:fig3} Schematics showing the transient heat transfer processes from a point heat source in He II. (a) At time $t$ less than the heat pulse duration $\Delta t$, a cavitation zone forms near the heater. (b) At $t>\Delta t$, some heat is carried out by the propagating second-sound zone.}
\end{figure}

\subsection{Transient heat transfer in He II}
It is known that heat transfer in He II is via a counterflow of the two fluid components, i.e., the normal fluid flowing away from the heat source carrying all the heat content while the superfluid moving in the opposite direct to compensate the fluid mass \cite{VanSciver2012Helium}. The heat flux $q$ is related to the normal fluid velocity $v_n$ as $q=\rho sTv_n$, where $\rho$ and $s$ are the helium density and specific entropy, respectively. When the heat flux is above a threshold of order 10$^{-2}$ W/cm$^2$ \cite{Vinen-PRS_1957}, quantized vortex lines are produced in the superfluid \cite{Vinen1957Mutual}, each carrying a single quantum of circulation $\kappa \approx 9.97\times 10^{-4}$ cm$^2$/s around its angstrom-sized core \cite{donnelly_book}. A mutual friction between the two fluids arises due to scattering of the thermal excitations off the vortices \cite{Vinen1957Mutual}. For a transient heat transfer from a point heat source in He II (see the schematics in Fig. \ref{fig:fig3}), the heat content is transported by a second-sound zone propagating at the speed $c_2$ \cite{VanSciver2012Helium}. The thickness of the second-sound zone is about $c_2\Delta t$, within which a counterflow can establish.

It is worthwhile noting that Shimazaki \emph{et al.} observed that in one-dimensional (1D) transient heat transfer of He II through a circular pipe, the injected heat is carried uniformly in the second-sound zone at low heat fluxes \cite{Shimazaki1995Second}. As the heat flux increases, the vortex-line density $L$ (i.e., vortex-line length per unit volume) increases. Above a threshold heat flux (i.e., about 5 W/cm$^2$ for $\Delta t$ of a few milliseconds), the interaction between the second-sound waves and the dense vortices can strongly distort the temperature profile in the second-sound zone, leading to the so-called ``limiting profile'' with the formation of a second-sound shock front \cite{Shimazaki1995Second, iida_visualization_1996}. In this situation, a significant fraction of the heat is converted to vortex energy, which slowly releases as the tangle decays. Nevertheless, as we shall show in Sec. \ref{sec:discussion}, the heat energy that goes to the vortices in our experiment is \emph{negligible}. This is because, unlike in the 1D heat transfer case, the heat flux drops rapidly away from a point heat source, and so does the vortex-line density.

Another process that is relevant to transient heat transfer in He II is film boiling at high heat fluxes \cite{VanSciver2012Helium}. The threshold heat flux for film boiling to occur in saturated He II depends on the duration of the applied heat pulse. For a heat pulse of a few milliseconds, this threshold is about 15 W/cm$^2$ \cite{Shimazaki1995Second, iida_visualization_1996,Hilton2005Direct}. In our experiment, the instantaneous heat flux from the heater surface is much higher than the film boiling threshold. Therefore, at $t<\Delta t$, a cavitation zone must form in the vicinity of the heater, as depicted in Fig. \ref{fig:fig3} (a). Outside this cavitation zone, some heat energy can be carried out by the propagating second-sound zone.

\subsection{Deformation of He$_2^*$ tracer lines}
To evaluate the deformation of a He$_2^*$ tracer line, let us consider a line segment at an initial distance $r_0$ from the miniature heater. As the second-sound shock front arrives at this line segment, it starts to move at the local normal fluid velocity $v_n$. If we assume that the heat transfer is isotropic towards all directions from the heater, $v_n$ is always along the radial direction and therefore the radial displacement of the line segment $dr$ in time $dt$ is given by:
\begin{equation}
dr=v_n\cdot dt=\frac{q(r)}{\rho sT}\cdot dt
\label{eq:dr}
\end{equation}
where $q(r)=\dot{Q}_s/(2\pi r^2)$ is the heat flux across the hemisphere of radius $r$, with $\dot{Q}_s$ being the instantaneous rate of heat transfer over the entire hemispherical surface. By integrating Eq.~\ref{eq:dr}, one can derive the final distance $r_f$ of the line segment from the heater as:
\begin{equation}
r_f^3=r_0^3+\frac{3}{2\pi\rho sT}\int_{t_0}^{t_0+\Delta t} \dot{Q}_s dt=r_0^3+\frac{3Q_s}{2\pi\rho sT}
\label{eq:r}
\end{equation}
where $t_0$ is the time that the second-sound zone first arrived at the line segment, and $Q_s$ denotes the total heat energy carried by the second-sound zone.

\begin{figure}
	\includegraphics[scale=0.37]{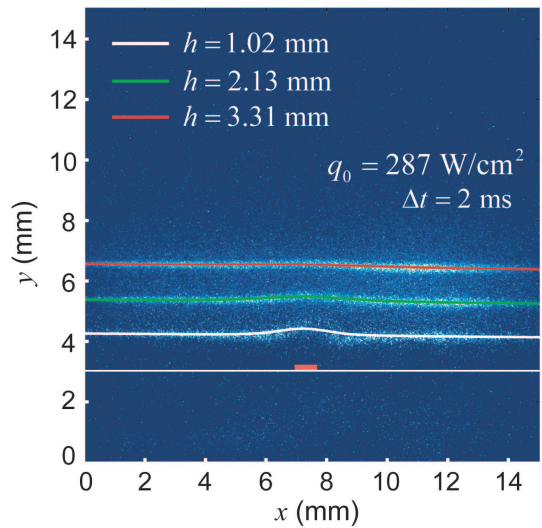}
	\caption{\label{fig:fig4}Examples of curve fittings to deformed tracer lines created at different initial height $h$ above the heater.}
\end{figure}

The assumption that the heat transfer is isotropic towards all directions holds true only if the heater size is small (i.e., approximately a point heat source) and that the effect of the PCB can be ignored. Note that due to the no-slip boundary condition of the normal fluid on the PCB, the heat flux within a boundary layer from the PCB surface must be different from that in bulk He II. The thickness of this boundary layer increases along the radius of the PCB and can be estimated to be about 300 $\mu$m near the edge of the PCB \cite{Hermann-book-1979}. Therefore, any relevant effect can be safely neglected.

Based on Eq. (\ref{eq:r}), the profile of a deformed tracer line can be computed from a given initial baseline, if we know the position $x_0$ of the heater on the PCB and the total heat $Q_s$ carried by the second-sound zone. In our analysis, we first adopt an algorithm developed by Pulkkinen \emph{et al.} \cite{Pulkkinen2014generative} to extract the locations of the baseline and the deformed line from the fluorescence images. We then set $x_0$ and $Q_s$ as two adjustable parameters to evolve the baseline profile so that a least squares fitting to the deformed line profile can be made. Typical examples of curve fittings to the deformed tracer lines based on their corresponding baselines are shown in Fig. \ref{fig:fig4}. One can see that this simple model very well reproduces the deformed line profiles. Fig. \ref{fig:fig4} also shows that the deformation of the tracer line becomes weaker when it is created at a larger distance $h$ from the heater. The minimum line deformation that can be resolved is comparable to about half the thickness of the tracer line. For a heat pulse of $q_0$=287 W/cm$^2$ and $\Delta t=2$ ms, we estimate that the maximum distance of the tracer line from the heater can reach $h_{max}$$\simeq$5 mm. Nevertheless, we note that based on Eq. (\ref{eq:r}), the displacement of the tracer line depends on the total heat transported through He II instead of the instantaneous heat flux on the heater surface. In a real quench event, the heat deposited in He II is comparable to the total energy stored in the cavity (i.e., of order 1-10 J \cite{padamsee_rf_2008}), which is two to three orders of magnitude larger than the heat produced by our miniature heater (i.e., of order 10 mJ). Therefore, we would expect resolvable line deformation even for tracer lines created at \emph{a few centimeters} away from the cavity surface.

\subsection{Analysis results}
\begin{figure}[tbp]
	\includegraphics[scale=0.37]{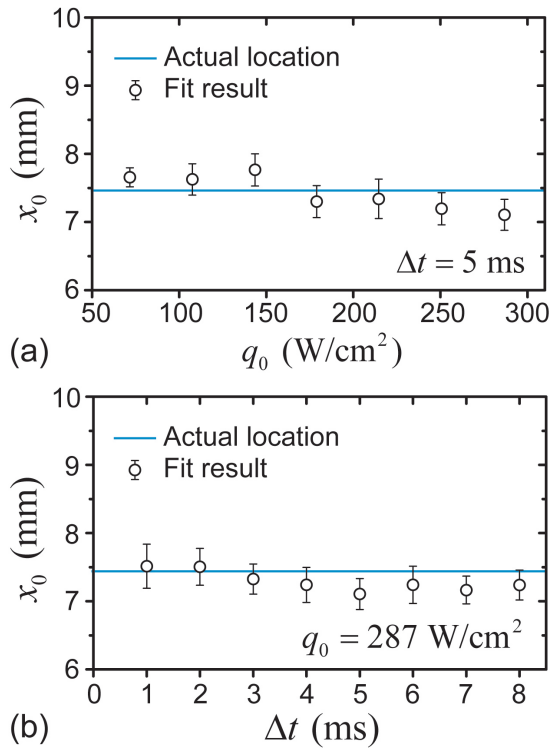}
	\caption{\label{fig:fig5} Fitting results of the heater location $x_0$ versus (a) heat flux $q_0$ at a fixed pulse duration, and (b) pulse duration $\Delta t$ at a fixed heat flux. $h$=1.98 mm for these measurements.}
\end{figure}

We have conducted the heater detection tests at various heat fluxes $q_0$=$\dot{Q}_0/A_h$ and pulse durations $\Delta t$. The values of $x_0$ and $Q_s$ can be determined through the curve fittings as we previously discussed. The results are collected in Table~\ref{tab:tab1}. Fig. \ref{fig:fig5} shows typical derived heater location $x_0$ in comparison with the actual center position of the heater (i.e., $x_h$=7.46 mm). For each test, we normally repeat our measurement 10 times so that the result uncertainty can be estimated. The fact that the obtained $x_0$ is always within a few hundred microns from the actual heater location, regardless of the applied heat flux $q_0$ and the pulse duration $\Delta t$, clearly proves the feasibility of this visualization-based non-contacting quench spot detection technology.

\begin{figure}[tbp]
	\includegraphics[scale=0.37]{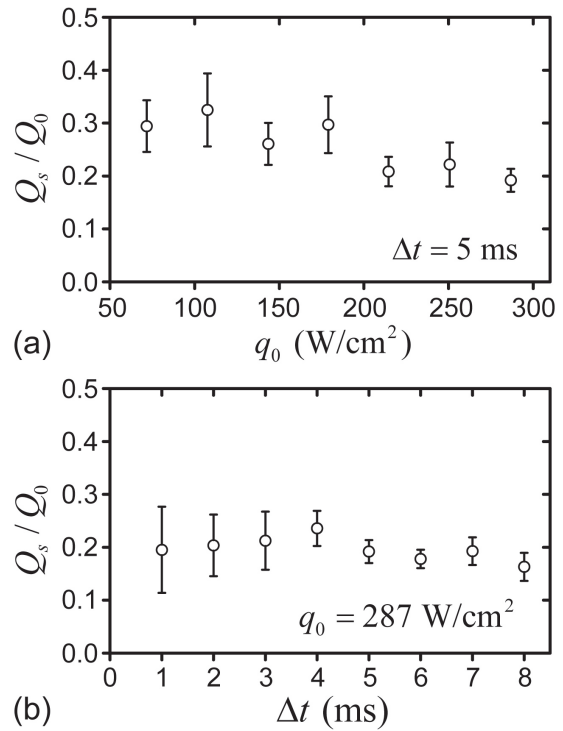}
	\caption{\label{fig:fig6} The ratio of the heat energy $Q_s$ carried by the second-sound zone to the total heat generated by the heater $Q_0$=$\dot{Q}_0\Delta t$ as a function of (a) heat flux $q_0$ and (b) pulse duration $\Delta t$. $h$=1.98 mm for these measurements.}
\end{figure}

In Fig. \ref{fig:fig6}, we show the ratio of the heat energy $Q_s$ carried by the second-sound zone to the total heat generated by the heater $Q_0=\dot{Q}_0\Delta t$ as a function of $q_0$ and $\Delta t$. This ratio $Q_s/Q_0$ appears to be weakly dependent on $q_0$ and nearly independent of $\Delta t$. The knowledge about the exact values of $Q_s$, which is not available from typical second-sound triangulation experiments, provides us a clue about the possible origin of the ``fast'' second sound evinced in those triangulation experiments (see detailed discussions in Sec. \ref{sec:discussion}).
From Table~\ref{tab:tab1}, we also note that under the same heater conditions, the fit values for $Q_s$ at $h$=1.02 mm appear to be much smaller than those obtained at larger $h$. This is likely due to the fact that the miniature heater can no longer be treated as a point heat source when the tracer line is placed too close.

\begin{table*}
	\caption{\label{tab:tab1}Results of the curve fittings at various experimental conditions.}
	\begin{ruledtabular}
		\begin{tabular}{ccccccccc}
			$h$&$q_0$&$\Delta t$&$t_d$&$x_0$&$Q_s$&$Q_s/Q_0$&$r_c$\\
			(mm)&(W/cm$^2$)&(ms)&(ms)&(mm)&(mJ)&&(mm)\\ \hline
1.02&	144&	2&	20&	7.64 $\pm$ 0.42&	0.17 $\pm$ 0.10& 	9.05\%&	0.30\\
1.02&	215&	2&	20&	7.30 $\pm$ 0.18& 	0.24 $\pm$ 0.06& 	8.35\%&	0.35\\
1.02&	287&	2&	20&	7.25 $\pm$ 0.13& 	0.29 $\pm$ 0.15& 	8.02\%&	0.39\\
1.98& 	78& 	5&	20&	7.66 $\pm$ 0.14& 	0.68 $\pm$ 0.11& 	29.43\%&	0.38\\
1.98& 	108& 	5&	20&	7.63 $\pm$ 0.23& 	1.12 $\pm$ 0.24& 	32.50\%&	0.49\\
1.98& 	144& 	5&	20&	7.77 $\pm$ 0.24& 	1.20 $\pm$ 0.18& 	26.07\%&	0.50\\
1.98& 	179& 	5&	20&	7.30 $\pm$ 0.23& 	1.70 $\pm$ 0.31& 	29.70\%&	0.60\\
1.98& 	215& 	5&	20&	7.34 $\pm$ 0.29& 	1.43 $\pm$ 0.19& 	20.85\%&	0.55\\
1.98& 	251& 	5&	20&	7.20 $\pm$ 0.24& 	1.78 $\pm$ 0.33& 	22.18\%&	0.61\\
1.98& 	287& 	1&	20&	7.51 $\pm$ 0.32& 	0.36 $\pm$ 0.15& 	19.52\%&	0.62\\
1.98& 	287& 	2&	20&	7.50 $\pm$ 0.27& 	0.75 $\pm$ 0.21& 	20.35\%&	0.63\\
1.98& 	287& 	3&	20&	7.33 $\pm$ 0.22& 	1.17 $\pm$ 0.30& 	21.25\%&	0.64\\
1.98& 	287& 	4&	20&	7.24 $\pm$ 0.26& 	1.73 $\pm$ 0.24& 	23.58\%&	0.68\\
1.98& 	287& 	5&	20&	7.11 $\pm$ 0.23& 	1.76 $\pm$ 0.20& 	19.20\%&	0.61\\
1.98& 	287& 	6&	20&	7.24 $\pm$ 0.27& 	1.96 $\pm$ 0.19& 	17.82\%&	0.59\\
1.98& 	287& 	7&	20&	7.16 $\pm$ 0.20& 	2.48 $\pm$ 0.34& 	19.27\%&	0.61\\
1.98& 	287& 	8&	20&	7.24 $\pm$ 0.22& 	2.40 $\pm$ 0.39& 	16.31\%&	0.56\\
2.13&	287&	2&	20&	7.46 $\pm$ 0.38& 	0.89 $\pm$ 0.21& 	24.42\%&	0.69\\
2.13&	287&	2&	30&	7.52 $\pm$ 0.18& 	0.98 $\pm$ 0.33& 	26.92\%&	0.72\\
2.13&	287&	2&	40&	7.38 $\pm$ 0.10&	0.93 $\pm$ 0.30& 	25.59\%&	0.70\\
3.31&	287&	2&	20&	7.05 $\pm$ 0.41& 	0.85 $\pm$ 0.19& 	23.30\%&	0.67\\
		\end{tabular}
	\end{ruledtabular}
\end{table*}

\section{\label{sec:discussion}Discussions}
\subsection{Partition of the heat energy}
\begin{figure}[tbp]
\includegraphics[scale=0.4]{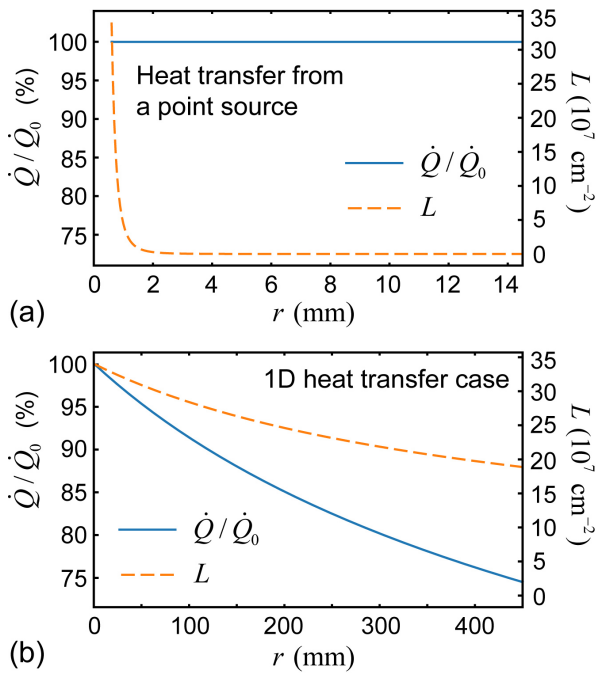}
\caption{\label{fig:fig7} Calculated vortex-line density $L$ and heat transfer rate $\dot{Q}=q(r)A$ at a distance $r$ from (a) a point heater such that $q_c$=15 W/cm$^2$ at $r_c$=0.6 mm; and b) a planar heater in a pipe that supplies the same heating power, i.e., $\dot{Q}_0=q_c\cdot 2\pi r_c^2$.}
\end{figure}

Since the heat $Q_s$ transported through He II by the second-sound zone is only a fraction of the total heat $Q_0$ generated by the heater, a natural question one may raise is: where does the remaining energy go? Indeed, a similar phenomenon was observed in earlier experiments on 1D transient heat transfer of He II through pipes \cite{Shimazaki1995Second, Hilton2005Direct, Zhang2004Use}. It was observed that the heat carried by the second-sound zone dropped significantly at heat fluxes greater than about 5 W/cm$^2$ from the heater surface. This observation was interpreted as due to the energy consumed in the formation of a dense vortex tangle in the pipe. The time evolution of the vortex-line density $L$ in a counterflow is governed by the so-called Vinen's equation \cite{Vinen1957Mutual}:
\begin{equation}
\frac{dL}{dt}=\alpha v_{ns} L^{3/2}-\beta \kappa L^2,
\label{eq:vinen}
\end{equation}
where $v_{ns}$=$(\rho/\rho_s)v_n$=$q/\rho_s sT$ is the relative velocity of the two fluids, and $\alpha$ and $\beta$ are dimensionless parameters with known values \cite{Donnelly-JPCRD_1998}. In a steady-state heat transfer, the equilibrium vortex density is given by $L_0=(\alpha/\beta \kappa)^2 v_{ns}^2$. The time $\tau$ taken for the line density to grow to the equilibrium value depends on the heat flux $q$ as $\tau$=$aq^{-n}$, with $a$ and $n$ being temperature-dependent constants \cite{Vinen-PRS_1957II, Shimazaki-Cryo_1998}. For a heat flux of order 10 W/cm$^2$, $\tau\simeq 0.3$ ms. Therefore, in transient heat transfer at high heat fluxes with a duration longer than 1 ms, it is reasonable to assume the equilibrium line density $L_0$ in relevant analysis. The energy $E$ associated with a random tangle of vortices per unit mass of He II is given by \cite{Vinen2002}:
\begin{equation}
E\approx\frac{\rho_s\kappa^2}{4\pi\rho}L\texttt{ln}\left(\frac{l}{\xi_0}\right),
\label{eq:energy}
\end{equation}
where $l=L^{-1/2}$ is the mean vortex-line separation distance and $\xi_0\simeq1${\AA} is the healing length for He II. Combining Eqs. (\ref{eq:vinen}) and (\ref{eq:energy}), one can derive an equation for the change in rate of the vortex energy $\dot{E}=\dot{E}_g-\dot{E}_d$, where the generation term $\dot{E}_g$ takes the form:
\begin{equation}
\dot{E}_g=\alpha v_{ns} L^{3/2}\frac{\rho_s\kappa^2}{4\pi\rho}\left[\texttt{ln}\left(\frac{l}{\xi_0}\right)-0.5\right].
\label{eq:eg}
\end{equation}
This generation term essentially accounts for the rate of energy that goes from the second-sound zone to the formation of vortices per unit mass of He II. Therefore, the heat flux $q(r)$ in the second-sound zone must satisfy:
\begin{equation}
\frac{d\left[q(r)A\right]}{dr}=-\dot{E}_g A/\rho,
\label{eq:flux}
\end{equation}
where $A$ is the cross section area at $r$, i.e., $2\pi r^2$ in our experiment or a constant in those 1D heat transfer experiments. Eq. (\ref{eq:flux}) allows us to calculate the rate of the heat energy, i.e., $\dot{Q}=q(r)A$, that can be transported at a distance $r$ from the heat source. The calculation results for both our experiment and the 1D heat transfer case are shown Fig. \ref{fig:fig7}. It is clear that in our experiment, $L$ drops rapidly with $r$ since the heat flux $q$ essentially decreases as $1/r^2$. Therefore, the heat energy carried by the 2nd sound zone suffers no noticeable attenuation. On the other hand, in the 1D heat transfer case, the vortex density remains high even at tens of centimeters away from the planar heater due to the slowly varying heat flux $q$. Consequently, the heat energy carried by the second-sound zone is constantly converted to vortex energy. This vortex energy then decays into heat that slowly diffuses out toward all directions \cite{VanSciver2012Helium}, which causes a broad temperature rise following the second-sound zone, as observed in some 1D heat transfer experiments \cite{Shimazaki1995Second}.

Accepting the conclusion that the vortex effect is negligible in our experiment, the only other mechanism that can consume the heat energy is the formation of the cavitation zone. The heat energy deposited near the heater surface can vaporize the helium atoms and lead to the nucleation and growth of small vapor bubbles. Indeed, in 1D heat transfer experiments \cite{Shimazaki1995Second}, it is suggested that the fraction of the heat energy consumed by this process increases with increasing the heat flux $q_0$ from the heater surface and can reach about 50\% at $q_0$=40 W/cm$^2$. Considering the much higher heat fluxes from the miniature heater surface in our experiment, it may not be surprising to see that over 70\% of the heat energy goes to the vapor bubbles.

We may make an order of magnitude estimation of the growth rate of the vapor bubbles. Considering a hemispherical vapor bubble that sits on the surface of the heater, if we assume that the injected heat is all utilized to vaporize the helium atoms, the growth of the bubble radius $R$ is then governed by \cite{Prosperetti-ARFM_2017}:
\begin{equation}
L_v\rho_v\frac{d}{dt}\left(\frac{2\pi}{3}R^3\right)=\pi R^2q_0,
\label{eq:bubble}
\end{equation}
where $L_v$ and $\rho_v$ are the helium latent heat and the vapor density in the bubble, respectively. According to Eq. (\ref{eq:bubble}), $\dot{R}=q_0/2L_v\rho_v$. For the heat flux used in our experiment or in typical cavity quenching (i.e., 10$^2$-10$^3$ W/cm$^2$), the bubble surface velocity $\dot{R}$ can exceed the first-sound speed in He II (i.e., c$_1$=230 m/s at 1.85 K \cite{Donnelly-JPCRD_1998}), which leads to the emission of strong first-sound shock waves due to the finite compressibility of He II. As a vapor bubble grows, the combined effects of buoyancy, shear lift, and contact pressure force may detach the bubble from the heater surface \cite{Gupta-AMR_2016}. Without the heat input, the bubble starts to shrink. Due to the existence of the heater surface nearby, the bubble collapses asymmetrically, leading to the formation of a micro-jet that can penetrate the bubble and impinge on the heater surface \cite{Lauterborn-JFM_1975, Vogel-JFM_1989, Liu-JTS_2013}. This process can again lead to strong first-sound emission. Therefore, the heat energy consumed by the vapor bubbles in the cavitation zone can essentially convert to acoustic energy carried by the first sound. Indeed, sound bursts and associated pressure spikes accompanying film boiling in He II have been observed experimentally \cite{Zhang2001Study, Bosque-ACE_2014}. Since the first-sound waves only cause the fluid parcels in He II to oscillate around their equilibrium positions, there is barely any detectable effect using either our flow visualization technique or those second-sound sensors.

\subsection{Possible origin of the ``fast'' second sound}
The conclusion that a large fraction of the heat energy is utilized in the creation of the cavitation zone has motivated us to propose a possible explanation for the seemingly fast second sound observed in many triangulation experiments. The formation of the cavitation zone is a very fast process, considering the rapid growth of the vapor bubbles as estimated in the previous section. The second-sound waves are indeed emitted from the surface of the cavitation zone instead of the heater surface. Therefore, for a second-sound sensor placed at a distance $S$ from the heater surface, the actual distance traveled by the second sound is $S'$=$S-r_c$, where $r_c$ denotes the size of the cavitation zone and is typically much smaller than $S$. As a consequence, the shorter travel time of the second-sound waves leads to a higher measured speed as given by $c_2'\simeq c_2(1+r_c/S)$. This simple idea is supported by the experimental observation that the fast second sound can be observed only in quench-spot experiments with high heat fluxes such that film boiling (i.e., cavitation) does occur near the hot spot \cite{liao_second_2012, junginger_high_2015}.

\begin{figure}[tbp]
\includegraphics[scale=0.4]{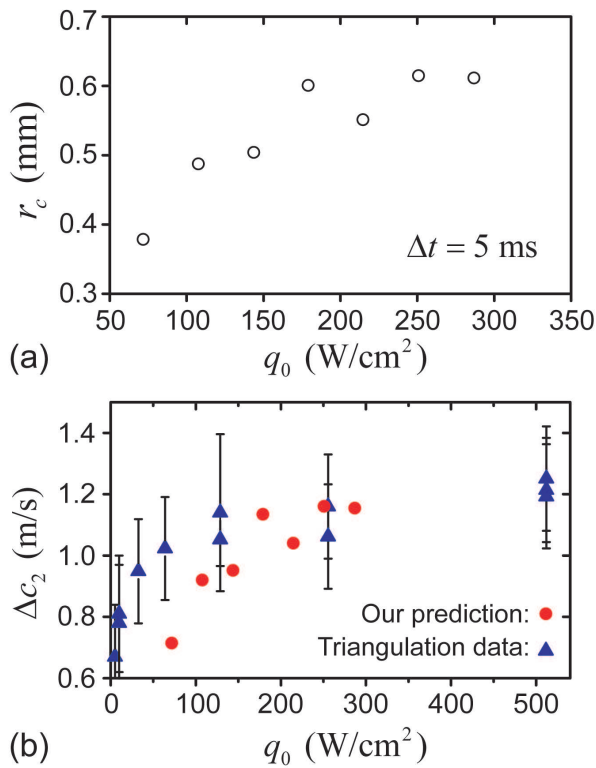}
\caption{\label{fig:fig8} (a) The estimated radius $r_c$ of the cavitation zone in our experiment as a function of the heat flux $q_0$. (b) Our predicted excess second-sound velocity $\Delta c_2$ in comparison with the triangulation data extracted from Refs.~\cite{junginger_high_2015, Koettig_IOP_2015}.}
\end{figure}

To evaluate the excess velocity $\Delta c_2$=$c_2'-c_2$, let us first estimate $r_c$. If we assume that the cavitation zone has a hemispherical surface with a radius $r_c$ and that the heat flux on this surface is about the threshold for film boiling (i.e., 15 W/cm$^2$ \cite{Shimazaki1995Second}), $r_c$ can be estimated based on:
\begin{equation}
\frac{\dot{Q}_s}{2\pi r_c^2}=\frac{Q_s/\Delta t}{2\pi r_c^2}=15~\texttt{W/cm}^2,
\label{eq:r_c}
\end{equation}
The calculated $r_c$ values for our experiment are listed in Table~\ref{tab:tab1}. $r_c$ appears to be nearly independent of the pulse duration $\Delta t$ but increases with increasing the heat flux $q_0$ from the heater surface as shown in Fig.~\ref{fig:fig8} (a). Since $\Delta c_2$=$(c_2/S)r_c$, this result agrees with the trend observed in previous triangulation experiments that the measured second-sound speed increases with increasing the heat flux \cite{junginger_high_2015, peters_advanced_2014}. More interestingly, since $r_c^2$ scales with $\dot{Q}_s$ (and hence $\dot{Q}_0$) according to Eq.~\ref{eq:r_c}, for a given heat flux $q_0$, $r_c^2$ then scales with the heater area $A_h$. If we consider the triangulation results reported in Refs.~\cite{junginger_high_2015, Koettig_IOP_2015} and scale $r_c^2$ based on their heater size $A_h$=15 mm$^2$, we can indeed compute the excess velocity $\Delta c_2$ for their OST sensor placed at $S$=5 cm from their heater surface. The results are shown in Fig. \ref{fig:fig8} (b). The blue triangles are triangulation data extracted from Fig. 5 in Ref.~\cite{Koettig_IOP_2015}. Amazingly, our predicted $\Delta c_2$ agrees quantitatively with the triangulation experimental observations. We should note that in reality the cavitation zone does not take a perfect hemispherical shape, especially when the heater size is large and the heat flux is small. This is probably why the triangulation data in Fig. \ref{fig:fig8} (b) show relatively large difference from our predictions at small heat fluxes. Nevertheless, the overall excellent agreement between our model prediction and the triangulation observations provides a strong support for the validity of our model.

\subsection{3D quench spot detection for real SRF cavities}
\begin{figure}[tbp]
	\includegraphics[scale=0.4]{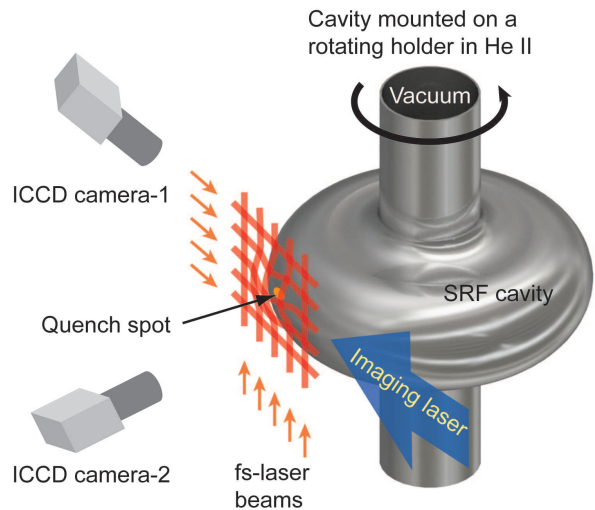}
	\caption{\label{fig:fig9} Schematic diagram showing the 3D quench spot detection scheme for real SRF cavities using a tracer-line grid.}
\end{figure}

In the proof-of-concept experiment presented in Sec. \ref{sec:exp}, we create the He$_2^*$ tracer lines in the vertical plane above a chosen miniature heater. The deformation of the tracer line only provides us the position information of the heater along the line where the vertical plane intersects with the PCB. In order for quench spot detection on the 2D surface of a real SRF cavity, our technique needs to be advanced. For instance, a simple extension of the current method could be to create two orthogonal tracer lines near the surface. The deformations of the two lines will then provide us complimentary information about the hot-spot location along two orthogonal directions. A more preferable and accurate detection scheme that we would like to propose is shown in Fig. \ref{fig:fig9}.

Instead of creating two tracer lines, we may first shape the fs-laser beam into a laser sheet and then pass it through a screen with parallel thin open slots to create an array of tracer lines. This can be done since the maximum pulse energy of our femtosecond laser (i.e., 4 mJ) is far greater than necessary for the creation of a single tracer line (i.e., 60 $\mu$J \cite{Gao2015Producing}). Overlapping two such tracer-line arrays can form a tracer-line grid, which has already been demonstrated in molecular tagging experiments in water \cite{Hu-MST_2006}. We may create such a tracer-line grid near a cavity surface and implement 3D imaging using two ICCD cameras placed at different angles \cite{Bohl-EF_2001}. Once a quench event is detected through monitoring the dissipation of the RF field in the cavity \cite{conway_defect_2010}, we can send in the imaging laser pulses to visualize the tracer-line grid. Due to the heat transfer from the quench spot to He II, a local deformation of the grid is expected. The analysis of this deformation will likely involve more fitting parameters. For instance, we may assume a Gaussian temperature profile in the hot area on the outer surface of the cavity. Then, besides the center position of the hot area and the transported heat in He II, other parameters such as the width of the Gaussian profile and the curvature radius of the surface may also be needed in the convolution of the initial grid profile to its final deformed profile. Finally, a scanning procedure may be implemented. By mounting the cavity on a rotating holder, we may use the same tracer-line grid to scan across the entire surface of the cavity so as to identify all surface defects.

\section{\label{sec:summary}Summary}
We have conducted a proof-of-concept experiment to demonstrate the feasibility of a flow-visualization based non-contacting technology for SRF cavity quench spot detection. By examining the deformation of a thin He$_2^*$ molecular tracer line created in He II nearby a miniature heater following a short heat pulse, we were able to reconstruct the heater location within a few hundred microns. The actual heat transported through He II by the propagating second-sound zone is found to be only a small fraction of the total injected heat energy. Our analysis shows that the remaining heat energy is essentially consumed in the formation of a cavitation zone surrounding the heater. The size of this cavitation zone is estimated based on the knowledge obtained about the transported heat. This information has allowed us to propose a new explanation for the decades-long puzzle observed in previous second-sound triangulation experiments regarding heat transfer at speeds higher than literature values. The excellent quantitative agreement between our predicted excess second-sound velocity and those measured in triangulation experiments provides a strong support of our model.

\begin{acknowledgments}
S.B. and W.G. acknowledge support from U.S. Department of Energy under Grant No. DE-FG02-96ER40952. The experiment was conducted at the National High Magnetic Field Laboratory, which is supported by National Science Foundation Cooperative Agreement No. DMR-1644779 and the State of Florida. The authors also wish to thank S. W. Van Sciver for valuable discussions and O. Yeung and M. Vanderlaan for their assistance in designing the helium cell.
\end{acknowledgments}

\bibliography{ref-extracts}

\end{document}